\documentclass[journal,twoside,final]{IEEEtran}
\usepackage{amsmath}
\usepackage{amssymb}
\usepackage{amsthm}
\usepackage{bm}
\usepackage{comment}
\usepackage{subfig}
\usepackage{float}
\usepackage{graphicx}
\usepackage{epstopdf}

\newtheorem{lemma}{Lemma}
\newtheorem{remark}{Remark}
\newtheorem{proposition}{Proposition}

\usepackage{cite}
\usepackage{enumitem}

\usepackage{xcolor}
\usepackage{lipsum}
\usepackage[ruled,norelsize]{algorithm2e}
\usepackage{algorithmic}

\usepackage{epstopdf} 

\allowdisplaybreaks

\begin{document}
%
\title{Fluid Aerial Networks: UAV Rotation for Inter-Cell Interference Mitigation
}

\author{Enzhi Zhou, Yue Xiao, Ziyue Liu, Sotiris A. Tegos,~\IEEEmembership{Senior Member,~IEEE,} \\ Panagiotis D. Diamantoulakis,~\IEEEmembership{Senior Member,~IEEE,} George K. Karagiannidis,~\IEEEmembership{Fellow,~IEEE}
\thanks{Enzhi Zhou is with the School of Computer and Software Engineering, Xihua University, Chengdu 610039, China (e-mail:ezzhou@mail.xhu.edu.cn).}
\thanks{Y Xiao is with the Provincial Key Laboratory of Information Coding and Transmission, Southwest Jiaotong University, Chengdu 610031, China (e-mail: alice\_xiaoyue@hotmail.com).} 
\thanks{Ziyue Liu is with the School of Aeronautics and Astronautics and Engineering Research Center of Intelligent Air-ground Integrated Vehicle and Traffic Control, Ministry of Education, Xihua University, Chengdu 610039, China (e-mail:liuziyue\_2006@126.com).}
\thanks{S. A. Tegos, P. D. Diamantoulakis and G. K. Karagiannidis are with the Department of Electrical and Computer Engineering, Aristotle University of Thessaloniki, 54124 Thessaloniki, Greece and with the Provincial Key Laboratory of Information Coding and Transmission, Southwest Jiaotong University, Chengdu 610031, China (e-mail: tegosoti@auth.gr, padiaman@auth.gr, geokarag@auth.gr).}
}

\maketitle

\begin{abstract}
With the rapid development of aerial infrastructure, unmanned aerial vehicles (UAVs) that function as aerial base stations (ABSs) extend terrestrial network services into the sky, enabling on-demand connectivity and enhancing emergency communication capabilities in cellular networks by leveraging the flexibility and mobility of UAVs.
In such a UAV-assisted network, this paper investigates position-based beamforming between ABSs and ground users (GUs).
To mitigate inter-cell interference, we
propose a novel fluid aerial network that leverages ABS rotation to increase multi-cell capacity and overall network efficiency.
Specifically, considering the line-of-sight channel model, the spatial beamforming weights are determined by the orientation angles of the GUs.
In this direction, we examine the beamforming gain of a two-dimensional multiple-input multiple-output (MIMO) array at various ground positions, revealing that ABS rotation significantly affects multi-user channel correlation and inter-cell interference.
Based on these findings, we propose an alternative low-complexity algorithm to design the optimal rotation angle for ABSs, aiming to reduce inter-cell interference and thus maximize the sum rate of multi-cell systems.
In simulations, exhaustive search serves as a benchmark to validate the optimization performance of the proposed sequential ABS rotation scheme. Moreover, simulation results demonstrate that, in interference-limited regions, the proposed ABS rotation paradigm can significantly reduce inter-cell interference in terrestrial networks and improve the multi-cell sum rate by approximately 10\% compared to fixed-direction ABSs without rotation.

\end{abstract}


\IEEEpeerreviewmaketitle

\section{Introduction}
Driven by the rapid advancement of aerial infrastructure and wireless technologies, the Internet of Things (IoT) paradigm is evolving from a purely terrestrial model to an integrated aerial-terrestrial framework. Unmanned aerial vehicles (UAVs), with their inherent mobility, altitude agility, and deployment flexibility, have emerged as key enablers of this transformation. UAVs support a broad spectrum of network functions, including operation as aerial base stations (ABSs), data relaying \cite{10504862}, data collection \cite{9989438}, edge computing \cite{10578306}, and autonomous networking \cite{10283826}.
When multiple UAVs are deployed collaboratively, they can form a flying ad hoc network, which serves as an aerial extension of conventional mobile ad hoc networks.
This evolution signifies a paradigm shift in next-generation wireless networks, transitioning from static ground-based architectures to dynamic three-dimensional (3D) aerial deployments.
In this context, UAV-assisted networks offer significant advantages in scenarios where traditional infrastructure is unavailable or requires rapid reinforcement by enabling location-specific, on-demand service provisioning with enhanced link quality, reduced path loss, and low interference.
Furthermore, this aerial connectivity paradigm provides connectivity from the sky, which is valuable for emergency response, coverage in remote areas, and temporary large-scale events. This positions UAVs as vital components of future intelligent, flexible, and resilient wireless ecosystems.

Among various UAV-enabled functions, an ABS, i.e., a UAV equipped with communication payloads and network control capabilities, serves as a flying antenna system that bridges the backhaul and access networks \cite{Zeng19}. Unlike passive relays or simple data collectors, ABSs operate as active network infrastructure, providing high-capacity, low-latency wireless links on demand. This makes them particularly effective in scenarios that require rapid network deployment, extended coverage, or enhanced service continuity.
Compared with static terrestrial BS, ABSs offer mobility, altitude adaptability, and flexible 3D placement, enabling agile and location-specific connectivity provisioning. These properties are especially beneficial in dynamic or underserved environments such as disaster response zones, rural areas, and temporary high-demand events \cite{Zeng19,Geraci22}. ABSs have thus emerged as a highly flexible and rapidly deployable solution to complement or temporarily replace ground-based communication infrastructure, addressing the growing need for resilient and adaptive wireless networks.

By adjusting their altitude and position in real time, ABSs can enhance line-of-sight (LoS) connectivity, improve coverage in obstructed or remote areas, and alleviate congestion during peak demand periods. Leveraging LoS-dominant air-to-ground channels, ABSs are well suited for operating in millimeter-wave (mmWave) frequency bands and supporting massive multiple-input multiple-output (MIMO)-based beamforming, enabling high data rate services for ground users (GUs) \cite{You20}. Under these conditions, beamforming vectors can be effectively computed using the positions of both UAVs and GUs \cite{Xiao22}. Furthermore, advances in lightweight communication payloads and autonomous flight control make ABSs increasingly compatible with LTE, 5G, and future 6G technologies.
However, integrating ABSs into existing networks introduces new challenges, including energy constraints, dynamic air-to-ground channel modeling, backhaul reliability, and severe inter-cell interference due to open-space propagation. These challenges call for intelligent network planning, resource allocation, and interference-aware beamforming strategies to ensure efficient and reliable UAV-assisted cellular communication systems.

\subsection{Literature Review}

Efficient and adaptive UAV-assisted communication requires research on ABS placement, which generally falls into three categories: classical single-objective optimization, joint optimization with communication parameters, and deployment strategies that consider practical constraints and real-world environments.

\subsubsection{Single-Objective UAV Placement}

Early studies primarily aimed to optimize UAV placement with respect to a specific objective, such as coverage, user satisfaction, or energy efficiency. A spiral-based deployment strategy was proposed in \cite{Lyu17} to ensure full GU coverage with the minimum number of ABSs. 
The work in \cite{Dinh19} shifted the focus to rate-constrained admission control, aiming to maximize the number of admitted users. Meanwhile, \cite{ChenRuirui20} proposed a density-boundary-prioritized strategy to enhance communication capacity per UAV. Energy-efficient placement under a given UAV budget was investigated in \cite{Li21} by jointly optimizing user association and power allocation. Furthermore, \cite{Guo22} considered both uplink and downlink transmissions to maximize the minimum average user rate, while \cite{Shabanighazikelayeh22} optimized 3D UAV positioning under altitude constraints and spatially varying user densities to reduce outage probability. Most recently, a multi-objective optimization framework was developed in \cite{Zhu24} to simultaneously maximize coverage utility and minimize total energy consumption, while accounting for the number of UAVs, 3D positioning, and mobility.

\subsubsection{Joint Optimization with Communication Parameters}

While the aforementioned works focused solely on placement, subsequent research has explored joint optimization frameworks, where UAV positioning is coupled with communication system design. In \cite{Alzenad18}, the vertical and horizontal components of single-UAV placement were decoupled to facilitate tractable optimization. 
The study in \cite{Xiao20} addressed beamforming optimization to maximize user sum rates under placement and constant modulus constraints. In \cite{Zeng24}, UAV placement and power control were jointly optimized to improve overall system performance. Content-aware deployment was considered in \cite{Luo22}, where UAV positioning was jointly optimized with caching strategies to minimize average content retrieval delay. Energy-aware UAV deployment for highway scenarios was addressed in \cite{Liao23}, incorporating user clustering and altitude control to increase uplink data rates. More recently, \cite{Liu24} investigated UAV-assisted integrated sensing and communication, jointly optimizing UAV locations and beamforming vectors to support dual functionalities.

\subsubsection{Placement under Practical Constraints and Intelligent Control}

To further enhance real-world applicability, several works have incorporated environmental factors and intelligent control techniques into UAV placement design. In \cite{Liu19}, coordinated multi-point  techniques were employed to mitigate inter-cell interference in multi-UAV networks. Post-deployment refinement to reduce interference was proposed in \cite{Liu21}, while reinforcement learning was used in \cite{Kaleem22} to jointly optimize UAV positions and transmit power in dynamic environments. Terrain-aware placement strategies were discussed in \cite{ChenJunting20, Li22}, where digital elevation and obstacle data were utilized. Statistical and map-based user distributions were leveraged in \cite{Wang22} and \cite{Bi24}, respectively, to support more realistic UAV placement. To model channel dynamics in urban settings, \cite{Yi24} introduced a blockage-aware air-to-ground channel model based on geographic information. Moreover, deep learning has been explored for data-driven placement optimization \cite{Wang24}, enabling UAVs to learn placement policies from spatial data. Considering beamforming characteristics, \cite{Miao20} derived beam weights based on user-UAV geometry, while \cite{Zhou25} modeled UAVs with directional antennas to more accurately characterize ground coverage.

\subsection{Motivation \& Contributions}

Although prior studies on ABS placement have considered inter-cell interference, most rely on simplified distance-based models that fail to capture the spatially varying interference patterns caused by directional beamforming of two-dimensional (2D) antenna arrays in practical UAV deployments.
The recently proposed six-dimensional movable antenna (6DMA) technology, which features distributed antennas capable of independent adjustment in both 3D position and orientation, overcomes the limitations of conventional fixed antennas by enabling base station antenna movement to dynamically optimize wireless channels between transmitters and receivers \cite{Shao25_6DMA, Shao25_6D}. Specifically, it introduces new degrees of freedom for enhancing the performance of multi-antenna systems, resulting in notable improvements in array gain, interference cancellation, and coverage in targeted areas \cite{Shao25Distributed, Zhu25Dynamic}.
Inspired by the same concept of 6DMA, {\color{black} UAVs } can be regarded as movable antenna arrays in the air, where the UAV orientation plays a critical role in shaping the multi-antenna channels between UAVs and GUs, particularly when 2D antenna arrays are employed. However, most existing UAV-enabled strategies focus solely on optimizing the spatial coordinates of UAVs, while overlooking the significant impact of UAV orientation on wireless channel characteristics.

Motivated by these insights, this work investigates how UAV rotation can be leveraged to manage inter-cell interference and enhance the performance of wireless cellular communication systems, and proposes a fluid aerial network architecture to enable flexible and adaptive interference-aware UAV deployment.
Although recent advances in environment-aware interference mitigation, such as reconfigurable intelligent surfaces  \cite{10974462} and fluid antenna systems  \cite{hong2025contemporarysurveyfluidantenna}, offer promising capabilities, these solutions typically require large-scale phase shift optimization or high-dimensional active beamforming. Such requirements lead to considerable computational complexity and impose significant hardware overhead.
In contrast, UAV rotation provides a lightweight and RF-chain-free alternative by physically adjusting the antenna array orientation, enabling interference management without additional signal processing or costly infrastructure modifications.
This strategy is particularly attractive for UAV platforms, where real-time beamforming is limited by the small number of onboard RF chains and constrained processing resources.
In this regard, we propose a UAV rotation-based strategy for practical and efficient interference mitigation in UAV-enabled communication systems.
Specifically, ABSs adopt maximum ratio transmission (MRT) beamforming based on the location information of GUs, thereby avoiding the need for complex multi-cell joint beamforming optimization. Furthermore, we analyze the beamforming gain patterns between UAVs and GUs under practical antenna configurations, and demonstrate that adjusting UAV orientation while maintaining fixed positions can effectively mitigate inter-cell interference.
In general, we propose a novel framework for optimizing UAV orientation to maximize the network sum rate through strategic UAV rotation. It should be noted that the UAV rotation is equivalent to rotating only the antenna array mounted on the UAV.

The main contributions of this work are summarized as follows:
\begin{itemize}
	\item  We model the position-based beamforming weights between ABSs and GUs under LoS channels, assuming that UAVs are equipped with 2D planar antenna arrays. By analyzing the positions of served and neighboring cell users, we investigate the beamforming gain of interference signals from UAV transmit beams affecting users in adjacent cells.
	\item We analyze the projection of beamforming gains onto the ground and demonstrate that adjusting the UAV orientation significantly influences the beamforming gain of interfering signals to users in neighboring cells, without affecting the serving cell's beamforming gain.
	\item  Based on these models, we formulate an optimization problem to maximize the system sum rate by rotating the UAVs. Moreover, to avoid the exhaustive search for the optimal rotation angles of multiple UAVs, we propose a sequential optimization algorithm that maximizes the current system sum rate and analyze its convergence and computational complexity.
	\item We simulate the multi-cell system to compare the system sum rate under fixed UAV orientations, exhaustive search, and the proposed sequential rotation scheme. The results show that adjusting UAV orientation can effectively reduce inter-cell interference and improve the average system sum rate by about $10\%$. Furthermore, the proposed low-complexity sequential UAV rotation scheme achieves performance similar to an exhaustive search.
\end{itemize}

\subsection{Structure} 
The rest of the paper is organized as follows. Section II presents the system model of multiple ABSs. Section III derives the interference beam power gain and demonstrates the effect of UAV rotation on the interference. Section IV formulates the system capacity maximization problem and proposes practical solutions. Section V presents the simulation results, where the system sum rate performance of the proposed scheme with different parameters is shown. Finally, Section VI concludes the paper.


\section{System Model}
We consider a multi-cell system assisted by UAVs, as illustrated in Fig. 1, where each UAV serves as ABS to provide downlink communication services to GUs within its corresponding cell, simultaneously serving $N_u$ GUs.
This architecture represents a fluid aerial network, in which the spatial configuration of UAVs can be dynamically adapted to the communication environment, enabling flexible and interference-aware service provisioning.

\begin{figure}
	\centering
	\includegraphics[width=1\columnwidth]{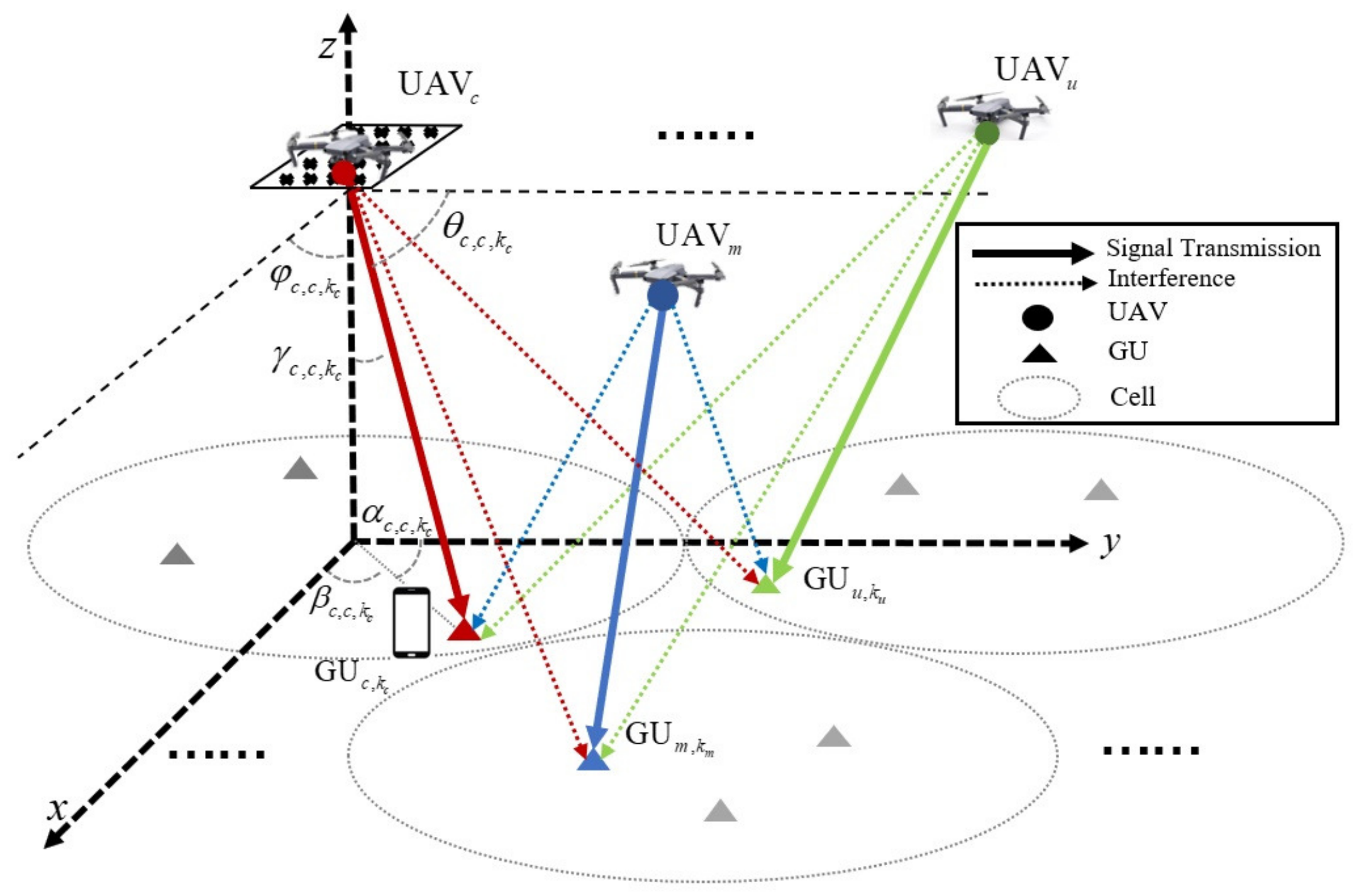}
	\caption{System model of a fluid aerial network for multi-cell communications.
    }\label{fig:system_model}
	\vspace{-0.1cm}
\end{figure}

Specifically, the spatial configuration of the UAVs and their associated GUs, along with potential inter-cell interference in the downlink, is also depicted in Fig. \ref{fig:system_model}. 
To support such air-to-ground communication, we assume that each UAV is equipped with a downward-facing square antenna array comprising $M \times M$ elements, while each GU is equipped with single antenna. 
We define the set of cells, the set of  GUs in the $c$-th cell, and the set of UAVs as $\mathcal{C}=\{1,...,N\}$, $\mathcal{K}_c=\{1,...,K_c\}$, and $\mathcal{U}=\{1,...,N\}$, respectively,  assuming that each cell is served by a single UAV.
For clarity, the ground is represented as the $x$-$y$ plane in a 3D Cartesian coordinate system, and the $u$-th UAV is assumed to operate at altitude $z_u$ above the ground. Accordingly, the location of the $k$-th GU in the $c$-th cell is denoted as $\bm{l}_{c,k_c} = \left( x_{c,k_c}, y_{c,k_c}, 0 \right)$, and the location of the $u$-th UAV is denoted as $\bm{l}_u = \left( x_u, y_u, z_u \right)$. 
Let $\mathbf{h}_{u,c,k_c} $ represent the direct channel between the $u$-th UAV and the $k_c$-th device in cell $c$. Specifically, the dominant signal transmission channel can be expressed by $\mathbf{h}_{c,c,k_c}$ and the interference channel is $\mathbf{h}_{u,c,k_c}$ with $u\neq c$. 

To uniformly describe all such channels, we define three angular parameters: $\theta_{.,c,k_c}$ and $\varphi_{.,c,k_c}$ as the azimuth angles with respect to the UAV's horizontal and vertical antenna arrays, respectively, and $\gamma_{.,c,k_c}$ as the pitch angle between the UAV and the GU. 
For notational simplicity, the subscripts are omitted in the following, i.e., the subscript of $\mathbf{h}_{.,c,k_c}$,  $\theta_{.,c,k_c}$, $\varphi_{.,c,k_c}$, and $\gamma_{.,c,k_c}$ are omitted for notational simplicity and reduced to $\mathbf{h}_{k}$, $\theta_{k}$, $\varphi_{k}$, and $\gamma_{k}$.
Accordingly, the steering vectors corresponding to the horizontal and vertical antenna arrays can be expressed, respectively, as follows: 
\begin{equation}
\begin{aligned}
{\bm{\psi}}_h \left( \theta_{k}  \right) &=  \frac{1}{\sqrt{M}}\left[ {1,{e^{j\pi \cos \left( \theta_k  \right)}}, \ldots ,{e^{j\left( {M - 1} \right)\pi \cos \left( \theta_k  \right)}}} \right],\\
{\bm{\psi}}_v \left( \varphi_k  \right) &= \frac{1}{\sqrt{M}}\left[ {1,{e^{j\pi \cos \left( \varphi_k  \right)}}, \ldots ,{e^{j\left( {M - 1} \right)\pi \cos \left( \varphi_k  \right)}}} \right].
\end{aligned}
\end{equation}

In addition, the system is considered in suburban or rural environments, where each UAV–GU link is dominated by a direct LoS channel due to negligible scattering.
As a result, the channel can be fully characterized by the steering vector between the UAV and the GU, which consists of both horizontal and vertical components and thus can be formulated as a vector in $1 \times M^2$ domain , i.e.,
\begin{equation}\label{eq:channel_model}
{\bm{h}}_k = a_k\bm{\psi}\left( \theta_k,\varphi_k\right) = a_k \bm{\psi}_h\left( \theta_k\right) \otimes \bm{\psi}_v\left( \varphi_k \right),  
\end{equation}
where $\otimes$ is the Kronecker product. $a_k$ is the channel phase depending on the signal transmission distance, normalized as ${{{\left| {a_k} \right|}^2}} = 1$.

To characterize the large scale fading dominated by the path loss, we first define the Euclidean distance between the $u$-th UAV and the $k_c$-th GU in the $c$-th cell as
\begin{equation}
\begin{aligned}
d_{u,c,k_c}  &= \| \bm{l}_u - \bm{l}_{c,k_c} \|_2\\
 &= \sqrt {{{\left( {x_u - x_{c,k_c}} \right)}^2} + {{\left( {y_u - y_{c,k_c}} \right)}^2} + {z_u^2}}.
\end{aligned}
\end{equation} 
To characterize the large-scale fading dominated by signal propagation path loss, the corresponding attenuation between the $i$-th  GU  and its serving UAV can be expressed as  
\begin{equation}
\begin{aligned}
    L_{u,c,k_c} & = {\left( {\frac{{4\pi d}}{\lambda }} \right)^2} \\ 
    & = \frac{{16{\pi ^2}}}{{{\lambda ^2}}} \left({{\left( {x_u - x_{c,k_c}} \right)}^2} + {{\left( {y_u - y_{c,k_c}} \right)}^2} + {z_u^2} \right) .
\end{aligned}
\end{equation}

{\color{black}
In this model, all GUs are assumed to be stationary or slowly moving, and those associated with the same UAV are scheduled on different time resources. 
In this system, due to the limited processing capability of the UAV, we consider that each UAV has only one RF chain, and thus the time-division single-user scheduling based on analog beamforming is adopted. Specifically, the UAV's antenna array adjusts the signal phase of each transmit antenna to direct the beam toward a target user within the cell during a given time slot. This scheme avoids complex beamforming algorithms, reduces beam switching frequency, while completely eliminating intra-cell user interference.}
Therefore, the primary interference affecting users originates from signals transmitted by neighboring cell UAVs and the signal received by the $k_c$-th GU at cell-$c$ can be expressed as
\begin{equation}\label{eq:received_signal}
\begin{aligned}
    {y_{c,k_c}} & = \sqrt {\frac{P}{{{L_{c,c,k_c}}}}} {{\mathbf{h}}_{c,c,k_c}}{\bf {x}}_{c,k_c} \\
    &\qquad + \underbrace{\sum\limits_{{u\in \mathcal{U}\backslash c}}    {\sqrt {\frac{P}{{{L_{u,c,k_c}}}}} {{\mathbf{h}}_{u,c,k_c}} {\bf {x}}_{u,k_u}  }}_{\text{inter-cell interference}}  + n,
\end{aligned}
\end{equation}
where $k_c\in \mathcal{K}_c$ and $k_u\in \mathcal{K}_u$. 
Moreover, the transmitted signal by UAV-$i$ to the $j$-th GU in cell-$i$ is denoted by 
\begin{equation}
\begin{aligned}
{\bf x}_{i,j}&={\bf f}_{i,j} { s}_{i,j},
\end{aligned}
\end{equation}
where ${\bf f}_{i,j} \in \mathbb{C}^{M^2\times 1 }$ is the  transmit beamforming vector.
${\bf s}_{i,j}$ is the corresponding transmitted symbol with normalized power. $P$ is the transmit power of UAVs and $n \sim \mathcal{N}\left( 0, \sigma^2_n\right) $ is the Gaussian white noise with standard deviation $\sigma_n$. 
Therefore, the interference received by the $k_c$-th GU in cell-$c$ is the sum of the signals transmitted by the other $N-1$ UAVs, excluding its dedicated serving UAV. 

It is also important to consider the role of the X2 interface in UAV-based cellular systems \cite{3gpp36243}. In the 3GPP New Radio (NR) architecture, X2 is a network-side interface that facilitates communication between base stations.
Functionally, it is dedicated to signaling coordination and control exchange, supporting inter-cell operations through the exchange of user scheduling information, interference measurements, and handover commands, these functions are essential for interference mitigation techniques such as inter-cell interference coordination, coordinated scheduling, and coordinated beamforming.
Therefore, the X2 interface can be realized via wireless backhaul links between neighboring ABSs in aforementioned UAV-assisted networks, enabling the exchange of position and control information. 
This inter-UAV communication capability forms the basis for implementing the proposed distributed interference-aware control algorithm. 
Based on above discussions, UAVs are assumed to obtain the location of its associated GUs, as well as the positions of nearby UAVs and GUs, through the X2 interface, which is sufficient for coordinating orientation and resource allocation, given that dominant interference primarily arises from adjacent cells.

Based on this system model, we further analyze the inter-cell interference caused by UAV transmissions in the following, evaluate the interference mitigation capability of UAV rotation, and develop a sum-rate maximization algorithm via orientation control.

\section{UAV Rotation-Assisted Beamforming for Interference Management}
In this section, we propose a practical beamforming and interference mitigation strategy for UAV-assisted communication systems. To reduce the real-time computational burden during flight and accommodate the limited number of RF chains on conventional UAV platforms, MRT is adopted as a low-complexity beamforming scheme.
Based on this MRT-based design, we first perform a 3D beamforming analysis to model the interference power from neighboring UAVs to GUs, taking into account realistic antenna configurations.
To further suppress inter-cell interference, we exploit UAV rotation as an additional controllable spatial degree of freedom. We then investigate the impact of UAV rotation angles on interference beam power gain, revealing a critical relationship between rotation geometry and interference intensity. Note that the antenna array mounted on the UAV can only be rotated via an appropriate gimbal mechanism instead of the entire UAV, resulting in the same behavior.

\subsection{Interference Gain Analysis}
Due to the broadcast of downlink transmission as shown in Fig. \ref{fig:system_model}, the beaforming from the $u$-th UAV toward the $k_u$-th GU at the cell $u$, i.e., GU-$\{u, k_u\}$ may cause interference to the $k_c$-th GU at the cell $c$ with $u\neq c$, then the corresponding interference gain observed at the GU-$\{c,k_c\}$ can be expressed by
 \begin{equation}\label{eq: interf_gain}
G_{c,k_c}=\sum \limits_{{u\in \mathcal{U}\backslash c}} \left| {\sqrt {\frac{P}{{{L_{u,c,k_c}}}}} {{\mathbf{h}}_{u,c,k_c}} {\bf {x}}_{u,k_u}  } \right| ^2.
\end{equation}
{\color{black} The interference power depends on the transmit power, path loss, and critically the beamforming gain in the direction to GU-$\{c,k_c\}$, which we denote as interference gain. }
Since the same transmit beam simultaneously acts as a desired signal for the intended user and a potential interference source for unintended users in neighboring cells, it is essential to analyze its spatial characteristics.

To facilitate practical onboard implementation, we adopt the MRT scheme at the UAV side. Specifically, the beamforming vector of UAV-$u$ intended for GU-$\{u, k_u\}$ is set as the conjugate transpose of the corresponding channel, i.e., ${\bf{f}}_{u,k_u} = \mathbf{h}_{u,u,k_u}^H$.
Under this design, the interference gain received by GU-$\{c, k_c\}$ from one UAV-$u$, which is dedicated to serving GU-$\{u, k_u\}$, can be expressed as
\begin{equation} \label{eq:interference_gain}
\begin{aligned}
g_{\{k_c,k_u\}}     &=    \left|{{\bf{h}}_{u,c,k_c}} {{\bf{f}}_{u,k_u}} s_{u,k_u}\right|^2 \\
                    &\overset{(a)}= |{\bf{h}}_{u,c,k_c} {{\bf{h}}_{u,u,k_u}^H}|^2.
\end{aligned}
\end{equation}
where step $(a)$ follows the MRT beamforming design and the normalized power of the signal.
In what follows, we simplify the notation by abbreviating the subscripts $\{u,c,k_c\}$ and $\{u,u,k_u\}$ as $\{k_c\}$ and $\{k_u\}$, respectively.
\begin{lemma}\label{lemma:interference_gain_decomposition}
The interference gain, as shown in \eqref{eq:interference_gain}, can be decomposed into horizontal and vertical components, as
 \begin{equation}\label{eq:expected_interf_gain}
g_{\{k_c,k_u\}} = \underbrace{\left| \bm{\psi}_h\left(\theta_{k_c}\right) \bm{\psi}_h\left(\theta_{k_u}\right)^H   \right|^2}_{{g}_h}  \underbrace{\left| \bm{\psi}_v\left(\varphi_{k_c}\right) \bm{\psi}_v\left(\varphi_{k_u}\right)^H \right| ^2}_{{g}_v} ,
\end{equation} 
 where $g_h$ and $g_v$ denote the interference gain in the horizontal and vertical domains, respectively.
\end{lemma}
\begin{IEEEproof}
 According to the channel model in \eqref{eq:channel_model}, the correlation between $\mathbf{h}_{k_u}$ and $\mathbf{h}_{k_c}$ can be expressed as
\begin{equation} \label{eq:channel_correlation}
\begin{aligned}
\bm{h}_{k_c}\bm{h}_{k_u}^H   &=   \left[ a_{k_c} \bm{\psi}\left( \theta_{k_c}, \varphi_{k_c} \right)\right]  \left[ a_{k_u}\bm{\psi}\left( \theta_{k_u}, \varphi_{k_u} \right) \right] ^H   \\
& = a_{k_c} a_{k_u}^H \left[ \bm{\psi}_h\left( \theta_{k_c}\right) \otimes \bm{\psi}_v\left( \varphi_{k_c}\right)\right] \\
&\qquad \qquad \qquad \times \left[ {\bm{\psi}}_h\left( \theta_{k_u} \right) \otimes {\bm{\psi}}_v\left( \varphi_{k_u}\right)\right]^H .
\end{aligned}
\end{equation}
According to the definition of the Kronecker product, we expand \eqref{eq:channel_correlation} by element-wise multiplication and obtain
\begin{equation}\label{eq:channel_correlation1}
\begin{aligned}
\bm{h}_{k_c}\bm{h}_{k_u}^H = a_{k_c} a_{k_u}^H \sum_{k=1}^{M} \sum_{l=1}^{M}  \psi_{k,h}\left( \theta_{k_c}\right)  \psi_{l,v}\left( \varphi_{k_c}\right) \\
\qquad \qquad \qquad \times  \psi_{k,h}^*\left( \theta_{k_u} \right)  \psi_{l,v}^* \left( \varphi_{k_u}\right),
\end{aligned}
\end{equation}
where $ \psi_{k,h} \left( \theta_{k_u} \right) $ and $ \psi_{l,v} \left( \varphi_{k_u} \right) $ denote the $k$-th element of $\bm{\psi}_h \left( \theta_{k_u} \right)$ and $l$-th element of $\bm{\psi}_v \left( \varphi_{k_u} \right)$, respectively. $\psi^*$ denotes the complex conjugate of $\psi$. As a result of expressing the channel components in the same direction as a vector inner product, \eqref{eq:channel_correlation1} can be rewritten as
\begin{equation}
\begin{aligned}
{\bf{h}}_{k_c}{\bf{h}}_{k_u}^H &= a_{k_c} a_{k_u}^H \sum_{k=1}^{M}  \psi_{k,h} \left( \theta_{k_c} \right)  \psi_{k,h}^*  \left( \theta_{k_u} \right) \\
&\qquad \qquad \qquad \times \sum_{l=1}^{M}  \psi_{l,v} \left( \varphi_{k_c}\right)  \psi_{l,v}^* \left( \varphi_{k_u} \right)\\
&= a_{k_c} a_{k_u}^H\left[ \bm{\psi}_h\left(\theta_{k_c}\right) \bm{\psi}_h\left(\theta_{k_u}\right)^H \right]  \left[ \bm{\psi}_v\left(\varphi_{k_c}\right) \bm{\psi}_v\left(\varphi_{k_u}\right)^H \right].
\end{aligned}
\end{equation}   
Hence, incorporating the normalized channel phase gain leads to \eqref{eq:expected_interf_gain}, which concludes the proof.
\end{IEEEproof}

The interference gain received by GU-\{$c,k_c$\} from the UAV-$u$ can be represented as a function of angles with the coordinates, as shown in the following proposition.
\begin{proposition}\label{prop:interf_gain_with_angles}
The interference gain received by GU-\{$c,k_c$\} from the UAV-$u$ is
\begin{equation}\label{eq:interference_beam_gain_expression}
\begin{aligned} 
 g_{\{k_c,k_u\}}\left( \Phi \right)  
 & = g_h\left( \Phi \right) g_v\left( \Phi \right)  \\&= \frac{{{{\sin }^2}\left( {\frac{{M\left( {\pi \cos {\alpha_{k_u}}\sin {\gamma_{k_u}} - \pi \cos {\alpha_{k_c}}\sin {\gamma_{k_c}}} \right)}}{2}} \right)}}{{M\sin^2 \left( {\frac{{\pi \cos {\alpha_{k_u}}\sin {\gamma_{k_u}} - \pi \cos {\alpha_{k_c}}\sin {\gamma_{k_c}}}}{2}} \right)}} \hfill \\ 
& \quad \times \frac{{{{\sin }^2}\left( {\frac{{M\left( {\pi \cos {\beta_{k_u}}\sin {\gamma_{k_u}} - \pi \cos {\beta_{k_c}}\sin {\gamma_{k_c}}} \right)}}{2}} \right)}}{{M\sin^2 \left( {\frac{{\pi \cos {\beta_{k_u}}\sin {\gamma_{k_u}} - \pi \cos {\beta_{k_c}}\sin {\gamma_{k_c}}}}{2}} \right)}}, 
\end{aligned}
\end{equation}
where $\Phi=\{{\alpha_{k_u}},{\beta_{k_u}},{\gamma_{k_u}},{\alpha_{k_c}},{\beta_{k_c}},{\gamma_{k_c}}\}$.
\end{proposition}
\begin{IEEEproof}
{\color{black}
 Based on Lemma 1, the interference gain can be decomposed into horizontal and vertical components, denoted by $g_h$ and $g_v$, respectively.
}
We then derive the interference gain of a linear array with the angles shown in Fig. \ref{fig:system_model}. Assuming that the antenna element spacing is half of the wavelength $\lambda$, the directional cosine of the angles $\theta$ and $\varphi$ can be represented by the angles $\alpha$, $\beta$ and $\gamma$, i.e.,
\begin{equation}\label{eq:direction_cosine}
\begin{gathered}
\Omega \left( \theta_i  \right) = \frac{{2\pi d\cos \theta_i }}{\lambda } = \pi \cos \theta_i = \pi \cos \alpha_i \sin \gamma_i \hfill \\
\Omega \left( \varphi_i  \right) = \frac{{2\pi d\cos \varphi_i }}{\lambda } = \pi \cos \varphi_i = \pi \cos \beta_i \sin \gamma_i. \hfill\\
\end{gathered} 
\end{equation}
where $i$ indicates $k_c$ or $k_u$. According to Fig. \ref{fig:system_model}, the ranges of angles are $0 \leqslant {\alpha_{k_c}},{\alpha_{k_u}} {\beta_{k_c}},{\beta_{k_u}} < \pi $ and $0 \leqslant {\gamma_i} < \frac{\pi }{2}$. 

Taking the horizontal direction as an example, the difference between the direction cosine of angles $\theta_{k_c}$ and $\theta_{k_u}$ is
\begin{equation}\label{eq:diff_directon_cosine}
\Delta \Omega\left(\theta_{k_c}, \theta_{k_u} \right)   = \Omega \left( {{\theta_{k_c}}} \right) - \Omega \left( {{\theta_{k_u}}} \right).
\end{equation}
The interference gain on the horizontal direction can be expressed by the directional cosine as
\begin{equation}\label{eq:expected_interf_gain_with_diff_direction_cosine}
\begin{aligned}
{{g_h }}  &= {\left| \bm{\psi}_h\left(\theta_{k_c}\right) \bm{\psi}_h\left(\theta_{k_u}\right)^H \right|^2} \\
&=\frac{1}{M}\sum\limits_{k = 1}^M \left| {{e^{j\left( {k - 1} \right)\pi \left( {\cos \left( {{\theta_{k_c}}} \right) - \cos \left( {{\theta_{k_u}}} \right)} \right)}}}\right| \\
&= \frac{{{{\sin }^2}\left( {M\Delta \Omega \left(  \theta_{k_c}, \theta_{k_u} \right)  /2} \right)}}{{M{{\sin }^2}\left( {\Delta \Omega \left(  \theta_{k_c}, \theta_{k_u} \right)  /2} \right)}}.\\ 
\end{aligned} 
\end{equation}
Similarly, the vertical interference gain $ {{g_v }} $ can also be obtained by replacing the angle $\theta$ in \eqref{eq:expected_interf_gain_with_diff_direction_cosine} with $\varphi$.
By combining  \eqref{eq:expected_interf_gain}, \eqref{eq:direction_cosine},  \eqref{eq:diff_directon_cosine}, and \eqref{eq:expected_interf_gain_with_diff_direction_cosine}, we can use the relative angles between the GUs and the UAV to represent the interference beam power gain as shown in \eqref{eq:interference_beam_gain_expression}, thus completes the proof.
\end{IEEEproof}
Therefore, by plugging \eqref{eq:interference_beam_gain_expression} into \eqref{eq: interf_gain}, the total interference can be further expressed as 
\begin{equation}\label{eq: total_interf_gain}
G_{c,k_c}=\sum \limits_{{u\in \mathcal{U}\backslash c}}  {  {\frac{P}{{{L_{u,c,k_c}}}}} g_{\{k_c,k_u\}}  },
\end{equation}
which can be used in the subsequent analysis to evaluate how UAV rotation influences inter-cell interference.

\subsection{Effect of UAV Rotation}
\begin{figure}
	\centering
	\includegraphics[width=0.84\columnwidth]{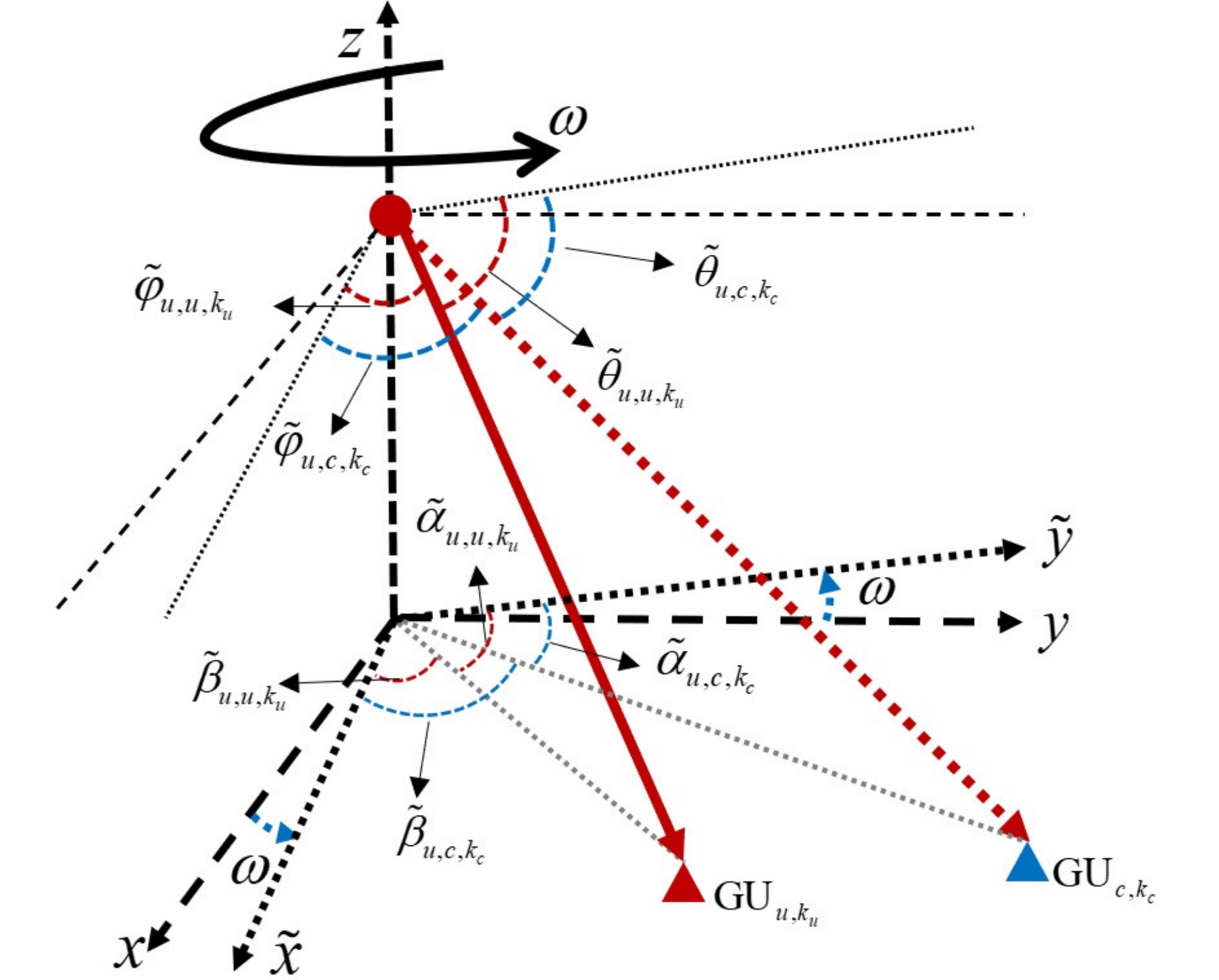}
	\caption{Illustration of the UAV rotation.}\label{fig:UAV_rotation}
	\vspace{-0.1cm}
\end{figure}

\subsubsection{Effect of UAV rotation on angular transformation}  
If the UAV rotates counterclockwise by an angle $\omega$, its orientation change can be equivalently modeled via a coordinate transformation. By aligning the system coordinate axes with the UAV's orientation, the relative angle variations between the UAV and each GU can be systematically captured. These variations, under the MRT scheme, form the basis for analyzing the spatial beam pattern during UAV rotation.

As shown in Fig. \ref{fig:UAV_rotation}, UAV rotation induces a variation in the relative angles $\alpha$ and $\beta$ between the GU and the coordinate origin, such that $\alpha$ increases while $\beta$ decreases accordingly, i.e.,
\begin{equation}\label{rotation}
\begin{aligned}
    &\tilde \alpha_i = \alpha_i + \omega, \\
    &\tilde \beta_i = \beta_i - \omega.
\end{aligned}
\end{equation}
where $i \in \{k_c, k_u\}$ represents the index of either the target or interfering ground user.
Note that the pitch angle $\gamma_i$ remains invariant under the UAV rotation. Following Proposition \ref{prop:interf_gain_with_angles}, the interference gain becomes a function of the rotation angle $\omega$, and is denoted as
\begin{equation}\label{eq:rotated_interference_beam_gain}
{\tilde g}_{ \{k_c,k_u\} }\left( \omega \right) = {g_{ \{k_c,k_u\} }}\left( \tilde \Phi \right).
\end{equation}
where $\tilde \Phi = \{{{\tilde \alpha }_{ku}},{{\tilde \beta }_{ku}},{\gamma_{ku}},{{\tilde \alpha }_{kc}},{{\tilde \beta }_{kc}},{\gamma_{kc}} \}$.
Note that, when simulating the UAV's orientation changes through coordinate system rotation, the spatial positions of the GUs undergo corresponding transformations in the new reference coordinate system, thus leading to variations in channel correlation and interference gain.

\subsubsection{Effect of UAV rotation on beam pattern}
\begin{figure}
	\vspace{-0.1cm}
	\centering
	\includegraphics[width=1\columnwidth]{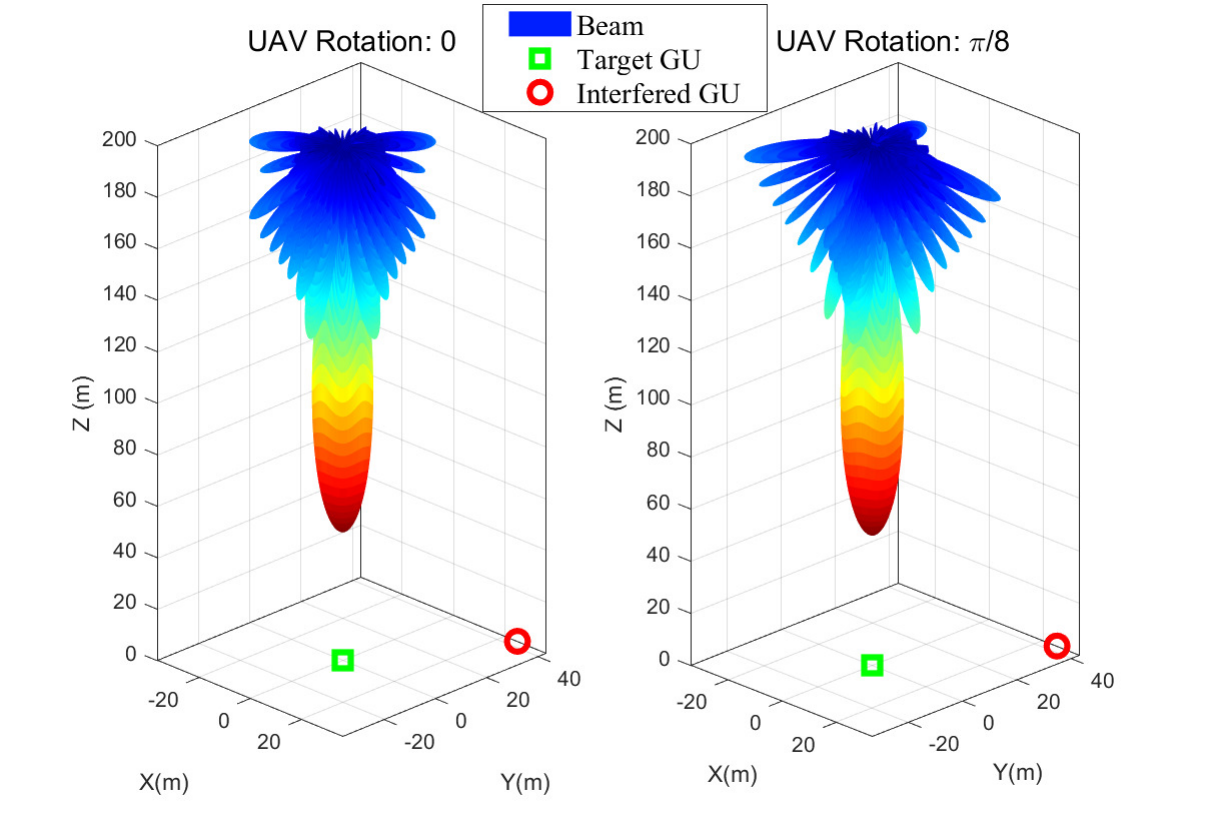}
	\caption{Evolution of the 3D beam pattern with UAV rotation.
}\label{fig:beam_3d_rotation}
	\vspace{-0.1cm}
\end{figure}

Beyond the geometric viewpoint, we explore how UAV rotation alters the interference pattern observed at GUs by adopting a beam gain perspective, especially with respect to main-lobe and side-lobe variations. Fig. \ref{fig:beam_3d_rotation} illustrates the evolution of the 3D beam pattern as the UAV rotates. When the UAV rotates from $0$ to $\pi/8$, the main lobe remains aligned with the intended target GU, while the side lobes rotate accordingly. Although the main lobe direction is unchanged, the side-lobe rotation can significantly affect the interference experienced by nearby users.  
This change in interference gain is primarily attributed to the variation in the angular position of the interfered GU relative to the UAV’s antenna array, which modifies the beamforming gain in that direction.
\begin{remark}
By altering the angular relationship between the antenna array and GUs, UAV rotation facilitates directional control of interference patterns without requiring complicated change in the beamforming weights. This property offers a low-complexity and hardware-efficient approach to interference mitigation in UAV-enabled wireless networks.
\end{remark}

\begin{figure*}
	\centering
	\hspace{-5pt}
	\subfloat[$\omega=0$.]{
		\label{fig:beam_gain_projection_1}
		\includegraphics[width=0.45\textwidth]{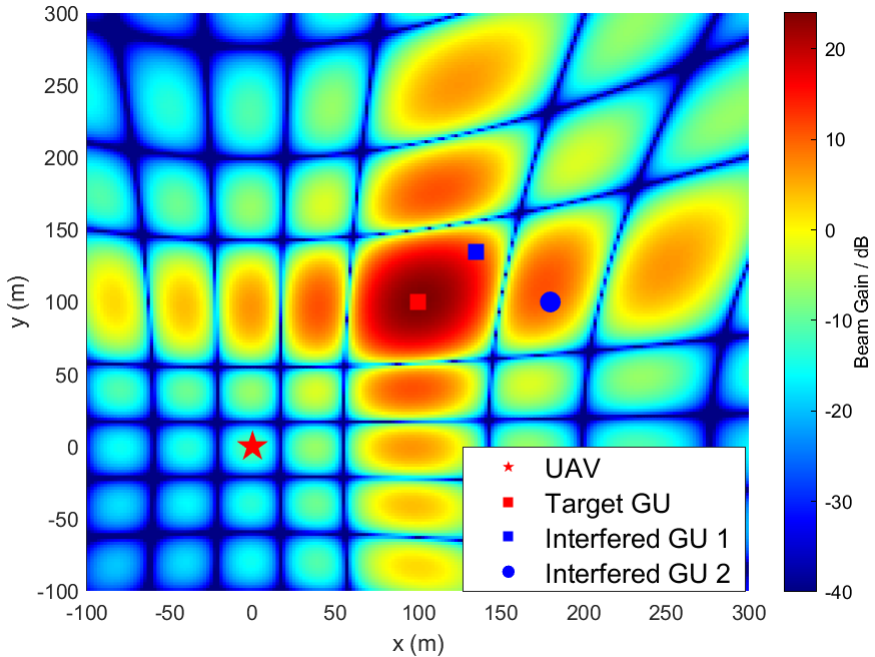}
	}\hspace{-5pt}
	\subfloat[$\omega=\frac{\pi}{8}$.]{
	\label{fig:beam_gain_projection_2}
	\includegraphics[width=0.45\textwidth]{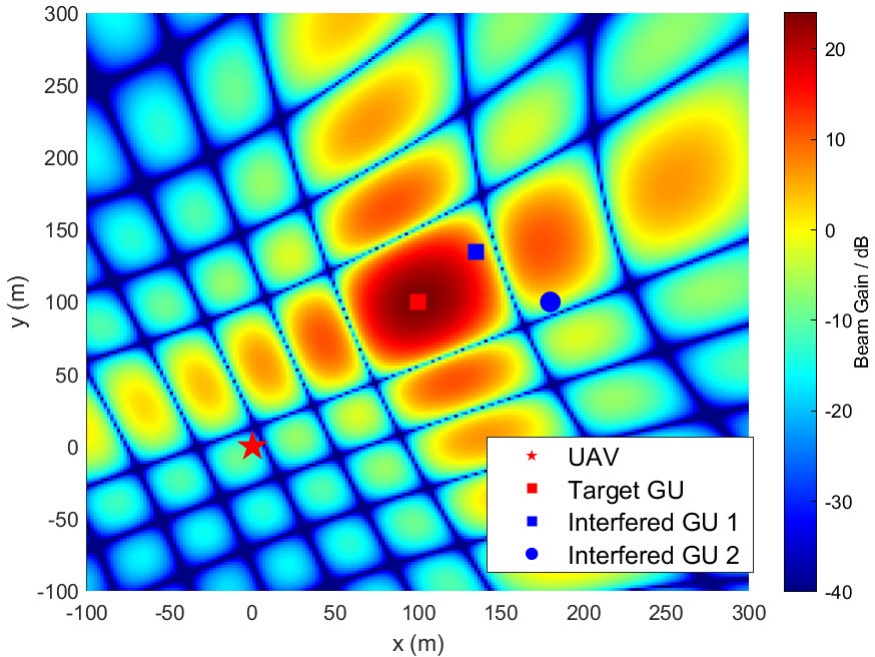}
	}\hspace{-5pt}
	\subfloat[$\omega=\frac{2\pi}{8}$.]{
	\label{fig:beam_gain_projection_3}
	\includegraphics[width=0.45\textwidth]{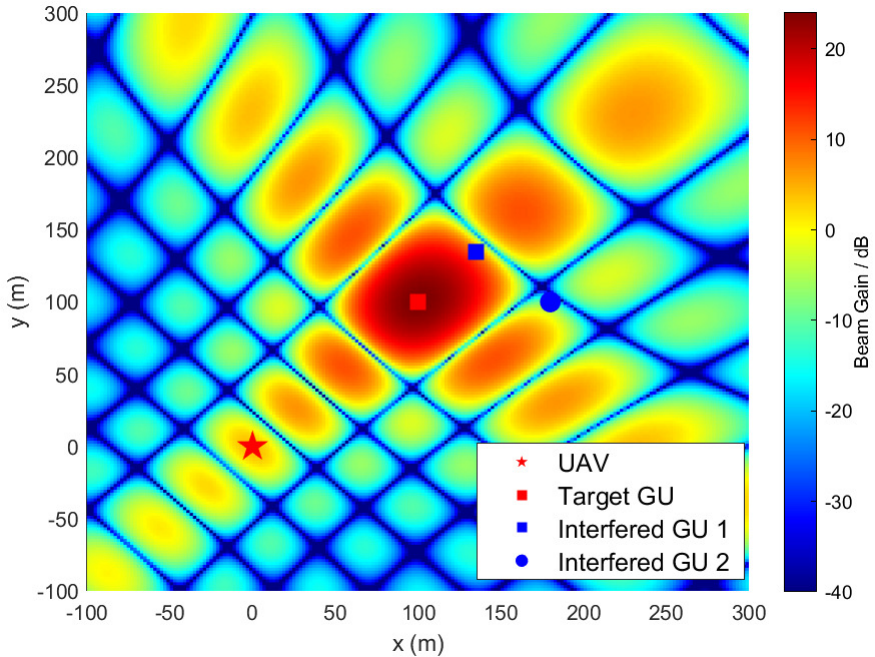}
	}\hspace{-5pt}
	\subfloat[$\omega=\frac{3\pi}{8}$.]{
	\label{fig:beam_gain_projection_4}
	\includegraphics[width=0.45\textwidth]{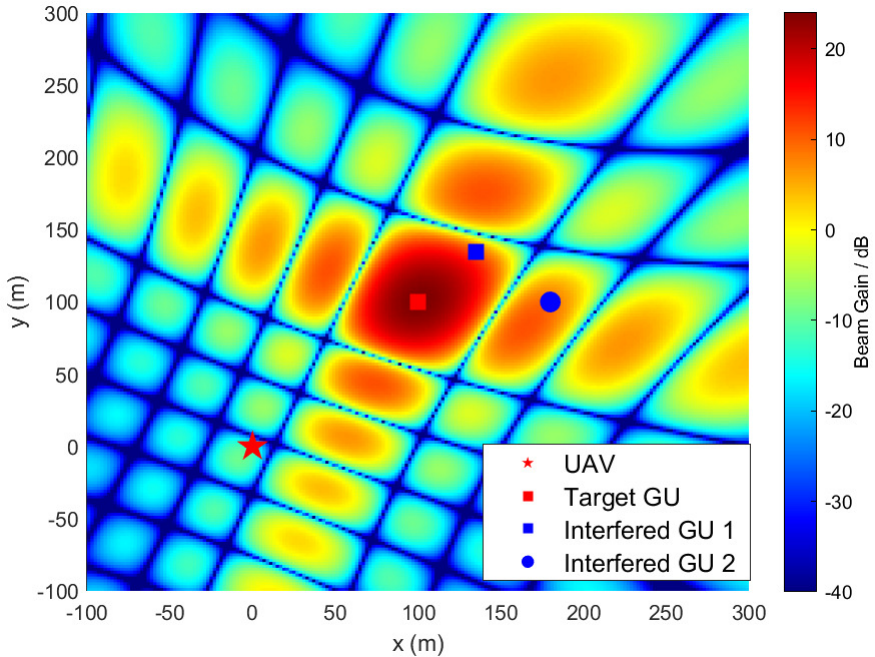}
	}
	\caption{The effects of UAV rotation on projected beam gain. 
    }\label{fig:interference_beam_gain_rotation}
	\end{figure*}

\begin{figure}[h]
	\vspace{-0.1cm}
	\centering
	\includegraphics[width=0.85\columnwidth]{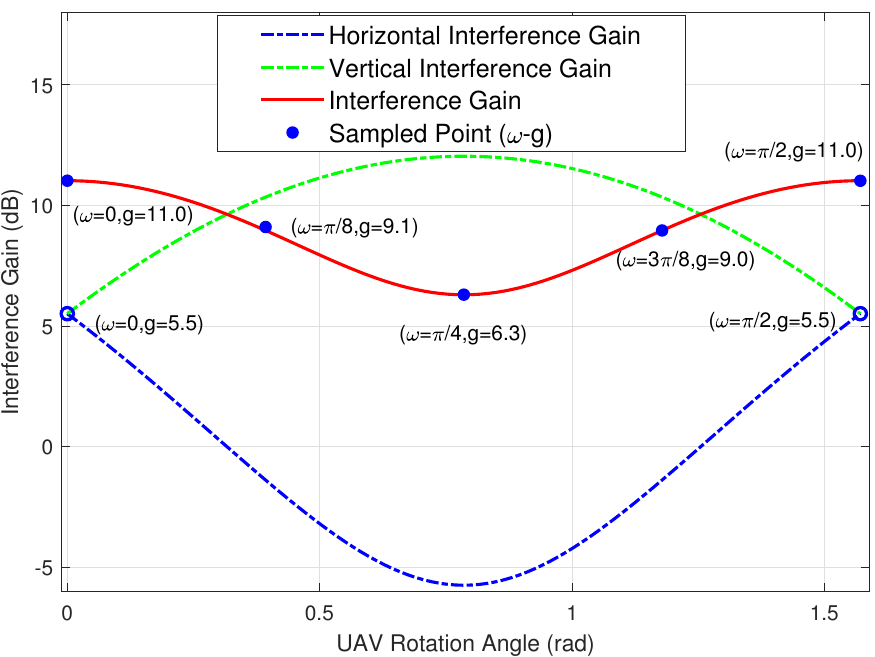}
	\caption{The effect of UAV rotation on the interference gain observed at GU1. }\label{fig:interference_beam_gain_rotation_135_135}
	\vspace{-0.1cm}
\end{figure}

\begin{figure}[h]
	\vspace{-0.1cm}
	\centering
	\includegraphics[width=0.85\columnwidth]{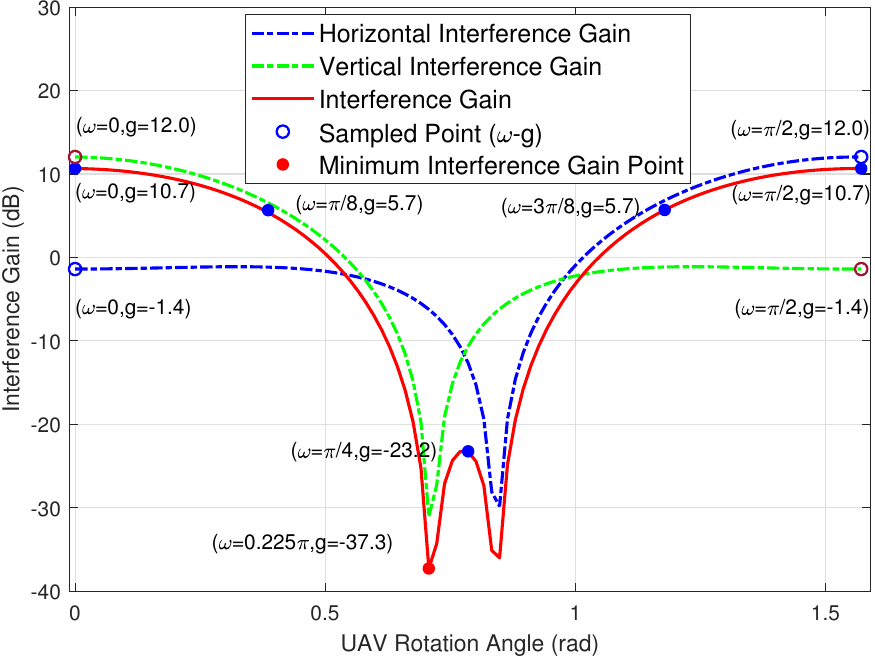}
	\caption{The effect of UAV rotation on the interference gain observed at GU2.}\label{fig:interference_beam_gain_rotation_180_100}
	\vspace{-0.1cm}
\end{figure}
\subsubsection{Effect of UAV rotation on interference mitigation}
Based on the above analysis of beam gain variation caused by UAV rotation, we further validate its effectiveness in mitigating interference at non-intended GU.

Fig.~\ref{fig:interference_beam_gain_rotation} illustrates the distribution of beam gains across multi cells when a $16\times16$-element square antenna array UAV is located at $(0, 0, 200)$m and steers its main beam toward the target GU at $(100, 100, 0)$m. Two interfered GUs are placed at $(140, 140, 0)$m and $(180, 100, 0)$m,  both experiencing interference from the UAV’s antenna array.
However, due to their distinct spatial positions, the interference gain observed at these users can be effectively mitigated through UAV rotation.
The UAV rotates counterclockwise from the x-axis by $\{0, \frac{\pi}{8}$, $\frac{2\pi}{8}$, $\frac{3\pi}{8}\}$ radians.
Figs.~\ref{fig:interference_beam_gain_rotation_135_135} and \ref{fig:interference_beam_gain_rotation_180_100} further show the interference beam power gains at these two interfered GUs across rotation angles.  

These figures reveal that the interference gain observed at the two interfered GUs exhibits distinct trends as the UAV rotates. For the user located within the main-lobe region, the gain decreases moderately, while for the user in the side-lobe region, the interference drops sharply due to side-lobe misalignment. Notably, the minimum interference occurs at specific rotation angles that depend on user positions.

\begin{lemma}
 
Due to the symmetric structure of the planar array, a UAV rotation of $\frac{\pi}{2}$ leads to an equivalent beam pattern as in the original orientation, resulting in the same interference gain observed at any GU.

\end{lemma}
\begin{IEEEproof}
Let $ i \in \{k_c, k_u\} $ denote the index of either the target or interfering GU. When the UAV rotates counterclockwise by $\omega = \frac{\pi}{2}$, the azimuthal angles transform as $\tilde \alpha_i = \alpha_i + \frac{\pi}{2}$ and $\tilde \beta_i = \beta_i + \frac{\pi}{2}$. 
Given the geometric constraint $\alpha_i + \beta_i = \frac{\pi}{2}$, we have
    \begin{equation}
    \begin{gathered}
        \cos {{\tilde \alpha }_i} = \cos \left( {\pi  - {\beta _i}} \right) = \cos {\beta _i}, \\
        \cos {{\tilde \beta }_i} = \cos \left( {\pi  - {\alpha _i}} \right) = \cos {\alpha _i}.
    \end{gathered} 
    \end{equation}
It follows that the horizontal and vertical channel correlation terms in Proposition~\ref{prop:interf_gain_with_angles}, as
    \begin{equation}
    \begin{gathered}
    {{g}_h}\left( \tilde \Phi \right) = {{g}_v}\left( \Phi  \right), \\
    {{g}_v}\left(\tilde \Phi\right) = {{g}_h}\left( \Phi  \right).
    \end{gathered} 
    \end{equation}
where $\Phi$ and $\tilde{\Phi}$ denote the angular parameter sets before and after rotation, respectively.

Therefore, the total interference gain remains unchanged under a \(\frac{\pi}{2}\) rotation:
 ${\tilde g}_{ \{k_c,k_u\} }\left( \frac{\pi}{2} \right) = {\tilde g}_{ \{k_c,k_u\} }\left( 0 \right)$, thus completes the proof.
\end{IEEEproof}
\begin{remark}
The $\frac{\pi}{2}$ periodicity of the interference pattern, as established in Lemma 2, allows the UAV rotation search space to be restricted to the interval $[0, \frac{\pi}{2})$. This constraint significantly reduces computational complexity and enables more efficient implementation of rotation-based interference mitigation algorithms.
\end{remark}
 \subsubsection{Practical utility of UAV rotation}
To quantify the spatial robustness of the UAV rotation angle, we define $\Delta\omega$ as the angular tolerance within which the interference remains minimal. A tangentially moving user must travel approximately $r \cdot \tan(\Delta\omega)$ to cause a significant angular deviation, where $r$ denotes the cell radius. 

As shown in Figs.~\ref{fig:interference_beam_gain_rotation_135_135} and~\ref{fig:interference_beam_gain_rotation_180_100}, the interference beam power gain decreases significantly when the UAV rotation angle lies between 0.6 and 0.8 radians, approaching its minimum, thus $\Delta\omega \approx 0.2$ radians is empirically determined based on the interference beam power gain curve, resulting in a displacement of about 100 m for a cell radius of 500 m, i.e., $500 \times \tan(0.2) \approx 101$ m. At a typical walking speed of 5 km/h, this corresponds to an update interval of approximately 60 seconds, indicating that UAV orientation can be adjusted at a relatively low frequency without compromising interference mitigation performance.

While this update interval may not appear long in absolute terms, it imposes minimal overhead in terms of control signaling and UAV actuation, and is well suited for low-mobility scenarios where user locations evolve gradually and predictably.

\begin{remark}
UAV rotation-based interference mitigation provides a lightweight physical-layer control mechanism that complements traditional beamforming strategies. By leveraging spatial robustness, the UAV can maintain effective interference suppression across a wide coverage region with minimal orientation updates. This approach is particularly well-suited for edge users and low-mobility scenarios, where user trajectories evolve slowly and predictably. Moreover, it enables efficient spatial interference shaping without requiring frequent beamforming reconfiguration or CSI feedback, thus reducing both computational and signaling overhead.
\end{remark}

\section{Fluid Aerial Network: Maximizing the System Sum Rate by UAV Rotation}

In this section, we employ the UAV rotation to a UAV-enabled multi-cell and multi-user cellular communication system with the objective of mitigating inter-cell interference and enhancing the overall system sum rate. In this direction, we first formulate a sum-rate maximization problem where the UAV rotation angles serve as optimization variables. Then, a low-complexity algorithm is proposed to determine the optimal orientations for each UAV.  


\subsection{System Sum Rate with UAV Rotation}

First, we analyze the average signal-to-interference-plus-noise ratio (SINR) at GU-$\{c,k_c\}$, which is served by UAV-$c$. We assume that each neighboring UAV $u \in \{\mathcal{U} \setminus {c}\}$ independently schedules one of its users $k_u \in \mathcal{K}_u$ in each time slot, whose transmission contributes to the interference observed at GU-$\{c,k_c\}$. Thus the resulting SINR is given by
\begin{equation}\label{SINR}
 \begin{gathered}
  {\eta _{c,{k_c}}} = \frac{{\frac{P}{{{L_{c,c,{k_c}}}}}{{\left| {{{\mathbf{h}}_{c,c,{k_c}}}{{\mathbf{f}}_{c,{k_c}}}} \right|}^2}}}{{\sum\limits_{{u\in \mathcal{U}\backslash c}} {\sum\limits_{{k_u} \in {\mathcal{K}_u}} {\frac{P}{{ {{K_u}} {L_{u,c,{k_c}}}}}{{\left| {{{\mathbf{h}}_{u,c,{k_c}}}{{\mathbf{f}}_{u,{k_u}}}} \right|}^2}} }  + \sigma _n^2}} \hfill \\
    \overset{(b)}= \frac{P}{{\sum\limits_{{u\in \mathcal{U}\backslash c}} {\sum\limits_{{k_u} \in {\mathcal{K}_u}} {\frac{P{L_{c,c,{k_c}}}}{{  {{K_u}} {L_{u,c,{k_c}}}}}{{\left| {{{\mathbf{h}}_{u,c,{k_c}}}{{\mathbf{f}}_{u,{k_u}}}} \right|}^2}} }  + {L_{c,c,{k_c}}}\sigma _n^2}} \hfill,
\end{gathered}
\end{equation}
where step (b) holds due to the use of normalized MRT beamforming, i.e., ${\left| \mathbf{h}_{c,c,{k_c}} \mathbf{f}_{c,k_c} \right|^2} = 1$.

Moreover, the received signal power of GU-$\{c,k_c\}$ is primarily determined by the UAV's transmit power and the corresponding path loss. As shown in \eqref{eq:interference_gain}, the interference from a neighboring UAV-$u$ is captured by the term $| \mathbf{h}_{u,c,k_c} \mathbf{f}_{u,k_u} |^2$, which depends on the relative azimuth and elevation angles between UAV-$u$ and GU-$\{c,k_c\}$. These angles are uniquely determined by the spatial positions of the UAV and the user, as well as the UAV’s orientation.
Given the positions of the GUs and UAVs, the relative angles of the GU-$\{c,k_c \}$ corresponding to the UAV-$u$ can be calculated by 
\begin{equation}\label{eq:angle_expression}
\begin{aligned}
{\alpha_{k_c}} &= \arccos \left( {\dfrac{{x_{c,k_c}}}{{\sqrt {{{\left( {x_{c,k_c} - x_{u}} \right)}^2} + {{\left( {y_{c,k_c} - y_{u}} \right)}^2}} }}} \right), \hfill \\
{\beta_{k_c}} &= \arccos \left( {\dfrac{{y_{c,k_c}}}{{\sqrt {{{\left( {x_{c,k_c} - x_{u}} \right)}^2} + {{\left( {y_{c,k_c} - y_{u}} \right)}^2}} }}} \right), \hfill \\
{\gamma_{k_c}} &= \arccos \left( {\dfrac{z_u}{{\sqrt {{{\left( {x_{c,k_c} - x_{u}} \right)}^2} + {{\left( {y_{c,k_c} - y_{u}} \right) }^2 + z_u^2}} }}} \right). \hfill
\end{aligned}
\end{equation}
Similarly, the angles ${\alpha_{k_u}}$, ${\beta_{k_u}}$, and ${\gamma_{k_u}}$ for GU-$\{u,k_u\}$ can be computed using its position in place of GU-$\{c,k_c\}$ in \eqref{eq:angle_expression}.

Considering the effect of UAV rotation, let $\omega_u$ denote the rotation angle of UAV-$u$. By substituting $\omega_u$ and the relative angles from \eqref{eq:angle_expression} into the rotation transformation in \eqref{rotation}, the resulting interference gain can be expressed as in \eqref{eq:rotated_interference_beam_gain}. Accordingly, the SINR expression in \eqref{SINR} can be reformulated to incorporate the rotation-dependent interference term. 
\begin{equation}
\begin{aligned}
    &\tilde \eta_{c,k_c}\left( \omega_1, \ldots,\omega_N\right) \hfill \\
    &= \frac{P}{{\sum\limits_{{u\in \mathcal{U}\backslash c}} {\sum\limits_{{k_u} \in {\mathcal{K}_u}} {\frac{P {L_{c,c,{k_c}}}}{{  {{K_u}} {L_{u,c,{k_c}}}}}{{\tilde g}_{\left\{ {{k_c},{k_u}} \right\}}}} } \left( {{\omega _u}} \right)   + {L_{c,c,{k_c}}}\sigma _n^2}}. \hfill \\
\end{aligned}
\end{equation}
Consequently, the achievable sum rate of the UAV orientation-aware system can be expressed as
{\small
\begin{equation}\label{sum_rate}
\begin{aligned}
&R\left(\omega_1,\ldots,\omega_{N} \right) \hfill \\
& =  \sum\limits_{c \in \mathcal{C}} \sum\limits_{k_c \in \mathcal{K}_c} {{{\log }_2}} \left( {1 + \frac{P}{{\tilde \eta_{c,k_c}\left( \omega_1, \ldots,\omega_N\right) + {L_{c,c, k_c}}}\sigma _n^2}} \right). \hfill \\
\end{aligned}
\end{equation}
}
This expression highlights that the system-wide sum rate is collectively determined by the rotation angles of all UAVs, as each orientation affects the interference experienced by neighboring users. Therefore, coordinated optimization of UAV orientations is essential for effective interference mitigation.

\subsection{Alternating UAV Rotation Scheme}
To mitigate inter-cell interference and improve system throughput, we formulate a rotation-aware sum-rate maximization problem by treating the rotation angles of all $N$ UAVs as optimization variables, i.e.,
\begin{equation}\label{eq:optimization_problem}
\begin{aligned} &\mathop {\arg\max }\limits_{\left\lbrace \omega_1,\ldots,\omega_{N} \right\rbrace }  R\left(\omega_1,\ldots,\omega_{N} \right) \hfill \\ &s.t. \quad {\omega_k} \in \left[ {0,\frac{\pi }{2}} \right],k = 1, \ldots ,{N}. \end{aligned} 
\end{equation} 
Due to the non-convex nature of the objective function, which involves intricate trigonometric and nonlinear couplings among the UAV rotation, gradient-based methods are inapplicable.
Although exhaustive search over all possible rotation combinations can yield the global optimum, it is computationally prohibitive when the number of UAVs or the angle resolution becomes large. 
To reduce the search complexity, the UAV rotation angle is confined to the range $\left[0, \frac{\pi}{2} \right]$, taking advantage of the array symmetry described in Lemma 2.

To make the optimization tractable, we introduce a zero-order coordinate-wise alternating method, referred to as alternating UAV rotation (AUR), which iteratively updates the rotation angle of each UAV while keeping the others unchanged.
The procedure is summarized in Algorithm \ref{alg:alternating_UAV_rotation_optimization}.

\begin{algorithm}
\caption{Alternating UAV Rotation (AUR) Optimization}
\label{alg:alternating_UAV_rotation_optimization}
\KwIn{Positions of ground users: $\{x_{c,k_c}, y_{c,k_c}\}$;  Positions of UAVs: $\{x_u, y_u, z_u\}$;  \qquad \qquad \qquad Angle discretizations resolution: $W$;
\qquad \qquad \qquad Max iterations $L$;
Convergence threshold $\epsilon$;}
\KwOut{Optimal UAV rotation angles $\{\omega_1^*,\ldots,\omega_{N}^*$\};}
\textbf{Initialize:} $\omega_u^{(0)} = 0$, $R^{(0)} = 0$, $\Delta R = \infty$, $l = 0$;
Pre-compute relative angles $\alpha$, $\beta$, and $\gamma$ using \eqref{eq:angle_expression};
\While{$l \leq L$ and $\Delta R > \epsilon$}{
$l = l+1$\;
		Update the optimal rotation angle of each UAV:\\
\For{$u = 1$ \KwTo $N$}{
\small{\small{
$\omega_u^{(l)} \leftarrow  \mathop {\arg \max }\limits_{{\omega_u}  = \left\lbrace 0,\frac{\pi }{{2W}}, \ldots ,\frac{\pi }{2} \right\rbrace} R(\omega_1^{(l)}, \dots, \omega_{u-1}^{(l)}, \omega, \omega_{u+1}^{(l-1)}, \dots, \omega_N^{(l-1)})$;
}
}}
$R^{(l)} \leftarrow R(\omega_1^{(l)}, \ldots, \omega_N^{(l)})$;
$\Delta R \leftarrow R^{(l)} - R^{(l-1)}$;
}
\end{algorithm}

The algorithm requires only position knowledge of UAVs and GUs and is well-suited for low-mobility or quasi-static deployments, where UAV orientation remains stable over extended durations.
Moreover, each iteration of AUR guarantees a non-decreasing sum rate and terminates when the improvement falls below a preset threshold $\epsilon$ or the maximum number of iterations is reached. The method yields a suboptimal but efficient solution with linear complexity in the number of UAVs and search granularity.

To adapt to the network connectivity of the ABSs, the execution mode of the proposed AUR algorithm supports either centralized or distributed implementation.
 \begin{itemize}
    \item \textbf{Centralized execution:}
  When ABSs are connected to terrestrial base stations via high-capacity backhaul links, the algorithm is centrally executed at the base station. The base station collects real-time location information of all UAVs and GUs, computes the optimal rotation angles for the UAVs, and broadcasts the results to the respective ABSs.
    \item \textbf{Decentralized execution:}
  When ABSs are connected to the core network via satellites, the algorithm is executed in a distributed manner due to the limited processing capability of satellite links. In this case, each UAV independently updates its rotation angle following a predefined update sequence determined by the core network. UAVs coordinate through a shared location dataset that contains the positions of all UAVs and their associated GUs. This dataset is progressively updated and forwarded along the sequence via X2 links. The first UAV monitors the dataset to detect any positional changes and triggers a new update cycle when needed. The last UAV terminates the algorithm once the system sum rate improvement $\Delta R$ falls below a predefined threshold $\epsilon$, or when the maximum number of iterations $L$ is reached.
 \end{itemize}
 
\begin{remark}
 For a system with $N$ UAVs and $W$ possible orientations per UAV, the computational complexity of an exhaustive search is $O(W^{N})$, which grows exponentially with the number of UAVs and quickly becomes impractical even for moderate-scale networks.
 In contrast, the proposed alternating optimization algorithm sequentially updates the orientation of each UAV and runs for at most $L$ iterations, resulting in a much lower complexity of $O(L N W)$. This linear scalability with respect to the number of UAVs enables efficient and real-time deployment in large-scale UAV systems with limited onboard computational resources.
 Moreover,  by adjusting the number of selectable angles $W$, the algorithm provides a flexible trade-off between computational cost and solution accuracy.
 \end{remark}

\subsection{Inaccurate GU Location Information}
In practical UAV-assisted networks, the effectiveness of X2-based inter-cell coordination can be significantly constrained by several factors, including the high mobility of GUs, the dynamic topology induced by UAV flight, and the latency or reliability limitations of the X2 interface particularly when UAVs rely on wireless backhaul connections.
As a result, UAVs may lack timely and accurate information about GUs in neighboring cells. This motivates the development of robust orientation optimization strategies that can tolerate inaccuracies in inter-cell GU location data.
To assess the robustness of the proposed algorithm, we evaluate the performance of the AUR algorithm under conditions where neighboring-cell GU positions are imperfectly known. 
Let the inaccurate coordinates of the GU-$\{c,c_k\}$ be modeled as
\begin{equation}
x_{c,k_c} = \tilde x_{c,k_c} + w_x, \quad y_{c,k_c}    = \tilde y_{c,k_c} + w_y,
\end{equation}
where $\tilde x_{c,k_c}$ and $\tilde y_{c,k_c}$ represent the real coordinates, and  $w_x~\text{and}~w_y\sim \mathcal{N}(0, \sigma_p^2)$ denote zero-mean Gaussian location errors with standard deviation $\sigma_p$.
These perturbed positions are then used in Algorithm \ref{alg:alternating_UAV_rotation_optimization} to determine the UAV orientations, thereby enabling performance evaluation under location uncertainty.

\section{Simulation Results and Discussions}

\begin{figure}[htbp]
	\centering
	\subfloat[UAV orientation angles $\left(0, 0, 0\right) $.]{
		\label{fig:interf_heatmap_fixed_direction}
		\includegraphics[width=1\linewidth]{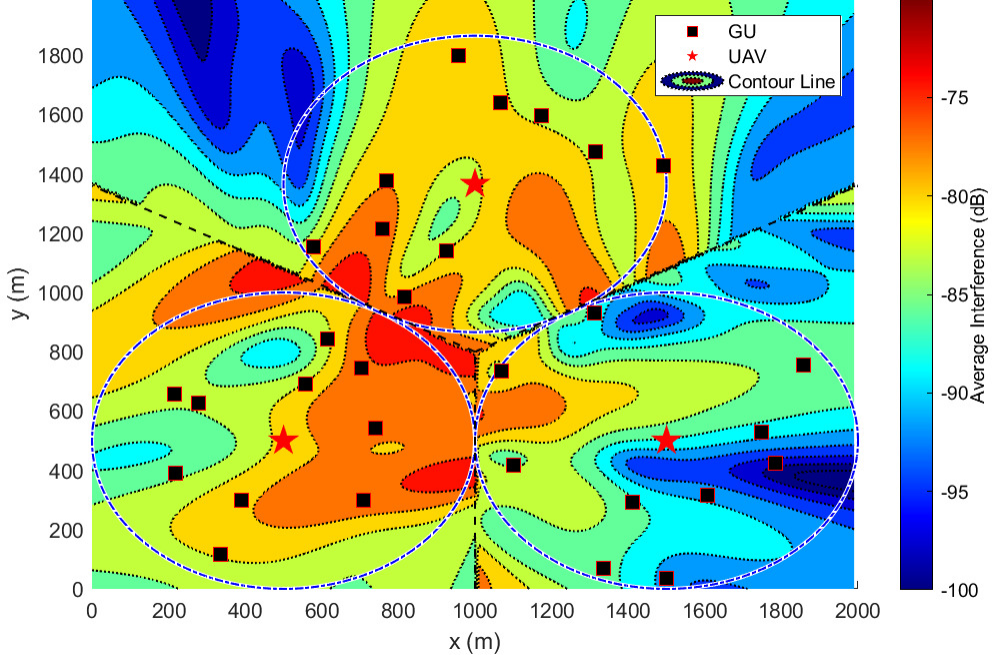}
	} \\
	\subfloat[UAV orientation angles $\left(0.44\pi, 0.28\pi, 0.19\pi\right) $.]{
		\label{fig:interf_dfheatmap_optimal_direction}
		\includegraphics[width=1\linewidth]{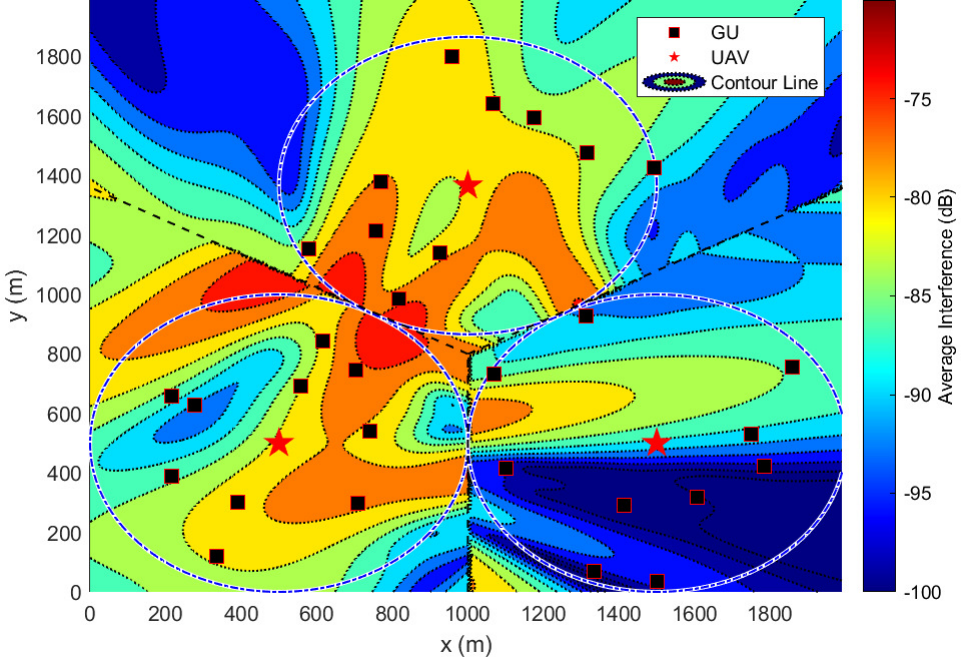}
   } 
	\caption{The inter-cell interference with UAV rotation, $M=8$.
    }\label{fig:interference_heatmap}
\end{figure}

This section presents a comprehensive performance evaluation of the proposed AUR optimization algorithm through systematic simulations. 
The main simulation parameters are detailed in Table~\ref{tab:simulation_parameter}.
We consider a typical deployment scenario where three UAVs are strategically positioned in a regular cellular configuration at coordinates $\{500, 500, 200\}$m, $\{500, 1500, 200\}$m, and $\{1000, 1500, 200\}$m.
Moreover, GUs are randomly distributed within each cell according to a circular distribution centered at the corresponding UAV, with a coverage radius of 500 m. To accurately capture inter-cell interference effects in the simulation, we enforce a minimum separation distance of 200 m between each GU and its associated serving UAV. {\color{black} Moreover, to ensure a fair comparison across different numbers of GUs, the average GU rate is defined as
\begin{equation}
    \bar{R}=\frac{1}{\sum\limits_{c \in \mathcal{C}} {{K_c}} }R,
\end{equation}
where $R$ denotes the total sum rate across all cells as shown in \eqref{sum_rate}. This definition normalizes the total throughput by the number of users, while also preserving the original optimization variables as in \eqref{eq:optimization_problem}.  }

\begin{table}
    \caption{Simulation Parameters}
    \label{tab:simulation_parameter}
    \centering
    \begin{tabular}{lc}
        \hline
        Parameter & Value \\
        \hline
        Number of UAV cells & 3 \\
        UAV altitude & 200 m \\
        Antenna elements (per dimension) & 8, 16, 32 \\
        Number of GUs per cell & 10, 30 \\
        Transmit power & 50 dBm \\
        System bandwidth & 1 GHz \\
        Noise power spectral density & $-174$ dBm/Hz \\
        Monte Carlo trials & 50 \\
        \hline
    \end{tabular}
\end{table}

\begin{figure*}
    \centering
    \subfloat[$M=8$.]{
        \label{fig:average_rate_8antennas}
        \includegraphics[width=0.32\linewidth]{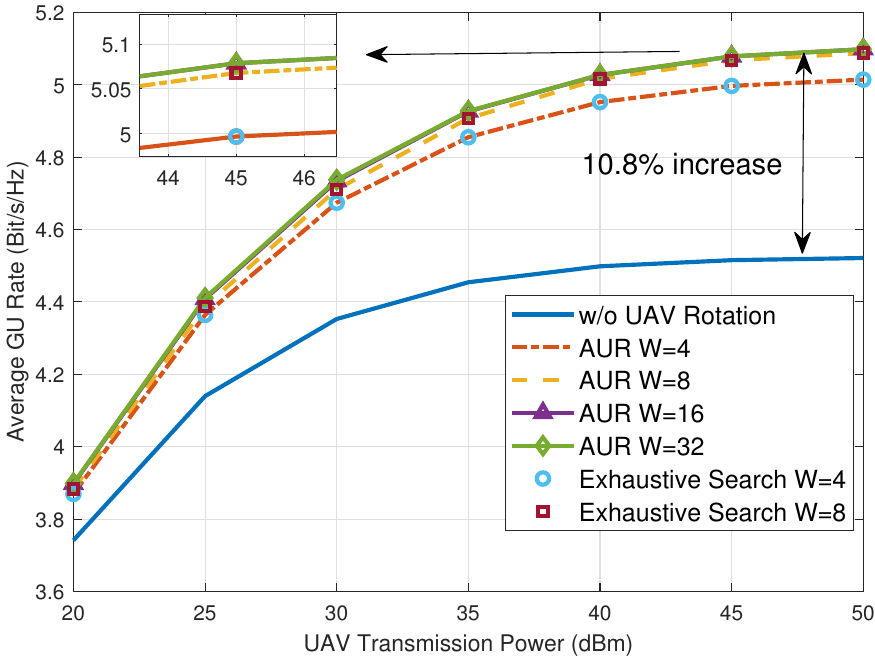}
    } 
    \subfloat[$M=16$.]{
        \label{fig:average_rate_16antennas}
        \includegraphics[width=0.32\linewidth]{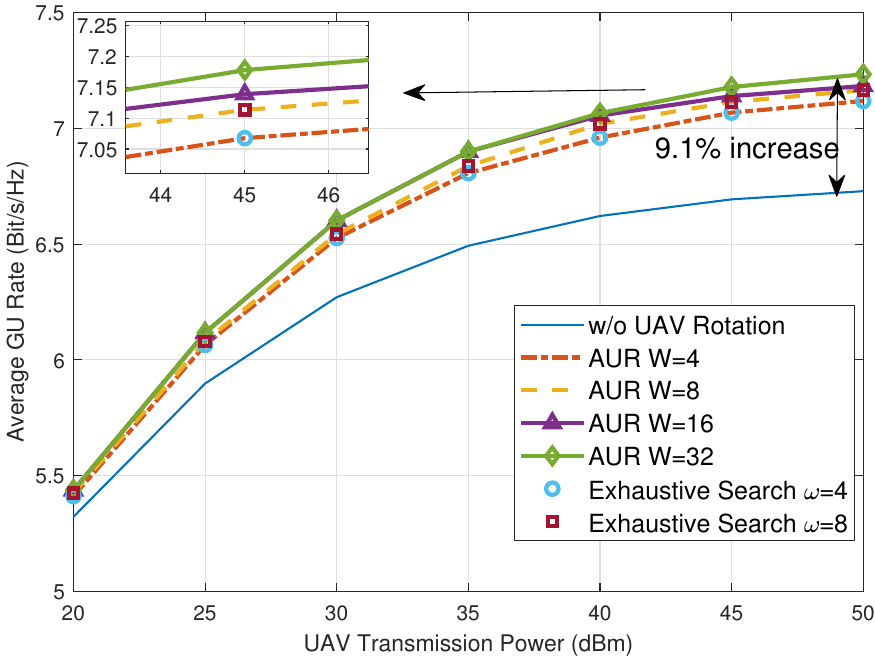}
    } 
    \subfloat[$M=32$.]{
        \label{fig:average_rate_32antennas}
        \includegraphics[width=0.32\linewidth]{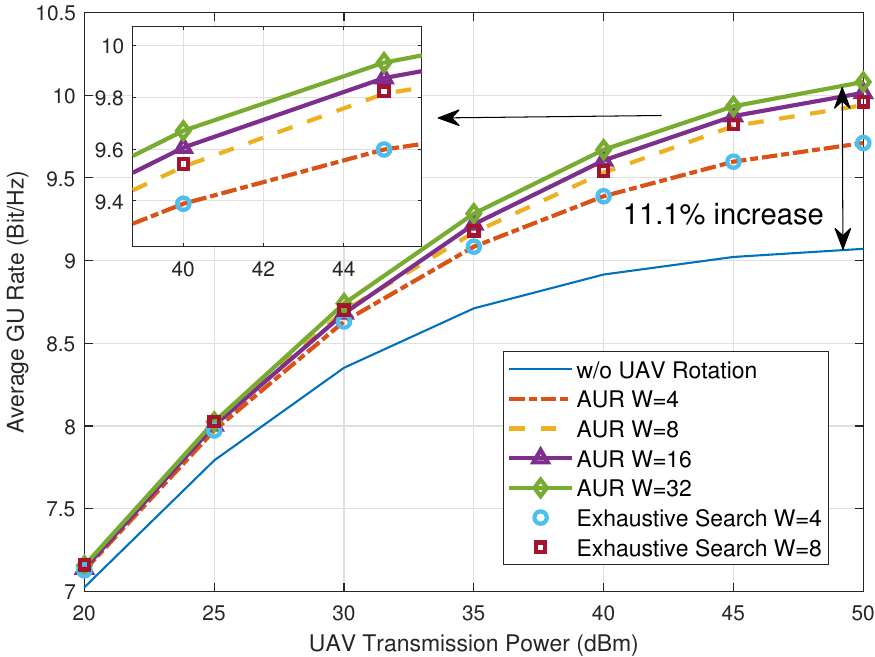}
    } 
    \caption{Average GU rate with the AUR algorithm under different numbers of UAV antennas.}
    \label{fig:average_rate_position}
\end{figure*}

 \begin{figure}
	\centering
	\includegraphics[width=0.85\columnwidth]{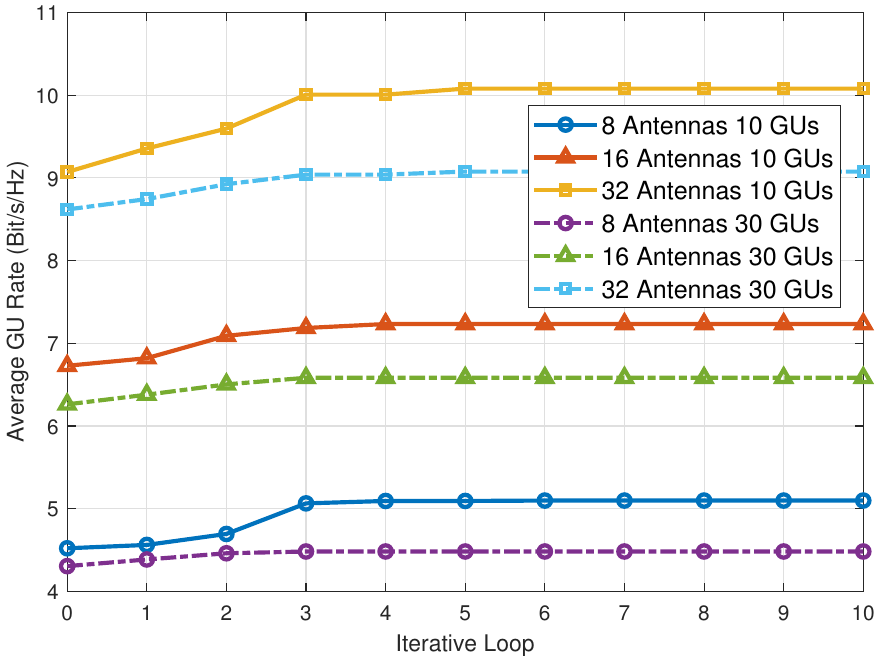}
	\caption{Average GU rate and convergence of AUR algorithm.} \label{fig:AUR_convergence}
\end{figure}

First, we provide an intuitive example to demonstrate how UAV rotation can change the average inter-cell interference power at different ground locations and improve the cell sum rate. In this example, each cell contains $10$ GUs located at the positions indicated by the black squares in the figure, while the UAV is positioned at the cell center, indicated by a red pentagram. Moreover, the UAV is equipped with an  $8\times8$ antenna array.
Fig.~\ref{fig:interference_heatmap} presents a heatmap of the average interference power received at various ground locations. Black dashed lines indicate the boundaries of the three UAV cells, while black dashed lines represent the user distribution regions within each cell. Notably, the interference observed at any location in a given cell originates from the UAVs in the other two cells.
Initially, the orientation angles of all three UAVs are set to $(0, 0, 0)$. After applying the proposed AUR algorithm, the optimal UAV orientation angles are determined to be $(0.44\pi, 0.28\pi, 0.19\pi)$. We compute and compare the average interference values under both the initial and optimized orientations. A comparison between Fig. \ref{fig:interf_heatmap_fixed_direction} and Fig. \ref{fig:interf_dfheatmap_optimal_direction} reveals that the proposed orientation adjustment significantly reduces the average interference power at several ground locations, thereby demonstrating the effectiveness of the AUR algorithm.

Next, we evaluate the average system sum rate under various configurations to demonstrate the effectiveness of the proposed AUR algorithm. In each scenario, three UAV cells are configured, and $50$ independent simulation trials are conducted. During each trial, the GUs are randomly distributed within the coverage area of each UAV cell. The baseline for comparison is the fixed UAV orientation with an absolute bearing angle of $0^\circ$.
To examine the impact of rotation angle granularity, we vary the number of discrete orientation angles $W$ as $4$, $8$, $16$, and $32$. Additionally, the proposed AUR algorithm is compared with the exhaustive search method, which is only feasible for $W = 4$ and $W = 8$ due to its exponential complexity.

Under these settings, Figs.~\ref{fig:average_rate_8antennas}–\ref{fig:average_rate_32antennas} illustrate the average system sum rate achieved by the proposed AUR algorithm under different numbers of antenna elements, i.e., $M = \{8, 16,32\}$. Compared to the fixed-orientation baseline, the AUR algorithm yields up to  $\{10.8\%, 9.1\%, 11.1 \% \}$ improvement in the high-SNR regime. For example, with $M = 8$ and $W = 32$, the average user rate increases from $4.55$ to $5.05$ bps/Hz. 
According to the Shannon capacity formula, this corresponds to an SINR increase of approximately $1.4$ dB. Similarly, the SINR gains for $M = \{16, 32\}$  are approximately $1.8$ dB and $3.1$ dB, respectively.
Notably, the AUR algorithm achieves performance comparable to exhaustive search while maintaining low complexity. When $M=8$, selecting $W=8$ directions captures most of the performance gain. For $M=\{16, 32\}$, a finer angular resolution ($W=32$) is required to fully exploit the narrower beamwidth, which enables more accurate interference steering but also increases computational cost.

In Fig. \ref{fig:AUR_convergence}, we evaluate the gain and convergence of the AUR algorithm under different antenna configurations and GU densities. The results show that the AUR algorithm converges to the optimal UAV orientation within $6$ iterations in all scenarios. The gain from UAV rotation increases with the number of UAV antennas, as narrower beams allow for more accurate inter-cell interference avoidance. However, as the number of GUs increases, the gain from UAV rotation decreases. This is because UAV rotation simultaneously affects the interference experienced by all GUs in adjacent cells. When GUs are distributed over many locations, reducing interference for all GUs simultaneously becomes impossible, resulting in reduced gains from UAV rotation. It can also be inferred that UAV rotation achieves greater gains when users are concentrated in one area.

\begin{figure}
	\centering
	\includegraphics[width=0.85\columnwidth]{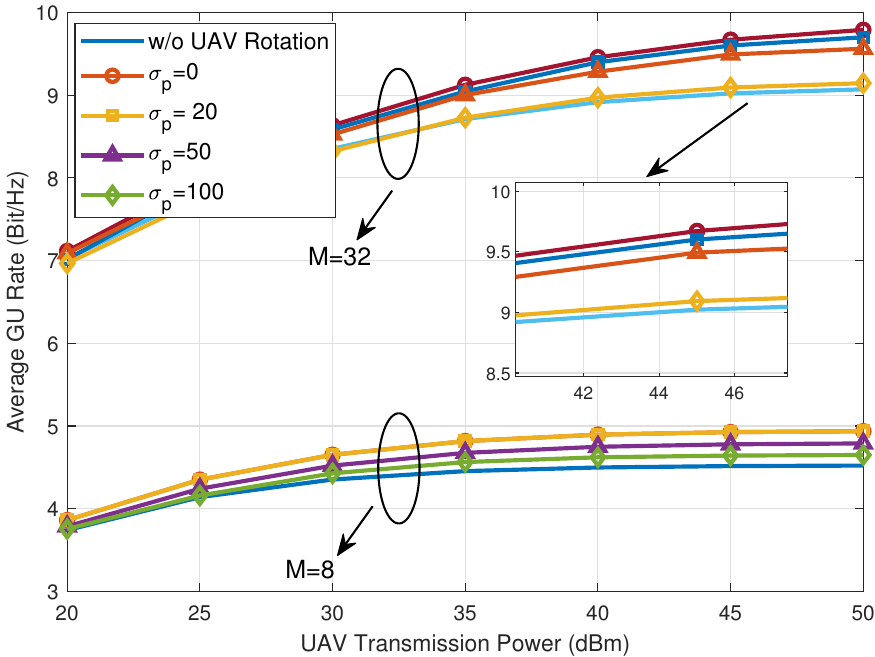}
	\caption{Average GU rate with inaccurate GU positions.} \label{fig:average_rate_position_error}
\end{figure}
 Fig. \ref{fig:average_rate_position_error} presents the average GU transmission rates under different positioning error standard deviations for UAVs equipped with $M=\{8, 32\}$, respectively.
The results demonstrate that the AUR gain gradually decreases as positioning errors increase. When the root mean square error reaches $20$ m, the AUR algorithm still achieves nearly full performance gain, indicating its inherent robustness against location inaccuracies. As the standard deviation of the positioning error further increases, the gain from UAV rotation diminishes. The results demonstrate that the robustness of the AUR algorithm is influenced by the number of UAV antennas. It show that when the root mean square error of the GU position reaches $100$ m, the UAV rotation can no longer achieve any performance gain. However, the $M=8$ configuration exhibits slow gain degradation, since fewer antennas result in wider beam projections on the ground, thereby relaxing the requirement for GU positioning accuracy.

\section{Conclusions}
This paper proposed a novel approach that leverages the mobility of UAVs in mmWave communications. Utilizing the characteristics of a 2D multi-antenna array, the rotation of deployed ABSs can reduce interference for users in neighboring cells without degrading performance for users in the local cell. We modeled the inter-cell interference experienced by GUs under UAV rotation and proposed a low-complexity multi-cell UAV rotation algorithm that significantly improves the sum rate of UAV-based multi-cell communication systems. The proposed scheme is broadly applicable because it mitigates inter-cell interference and has the potential to increase the capacity of multi-user systems and improve secrecy capacity for secure communications. Furthermore, it can be integrated with other techniques, such as UAV trajectory planning and user scheduling, to further improve the performance of UAV communication systems.

\vspace{-0.1cm}
\small



\end{document}